# Energy-Efficient Joint Optimization of VLC and Fog Computing Resources Allocation

Wafaa B. M. Fadlelmula, Sanaa Hamid Mohamed, Taisir E. H. El-Gorashi, Jaafar M. H. Elmirghani

*Abstract*—**In this paper, we consider an indoor networking architecture in which visible light communication (VLC) provides wireless connectivity between users and distributed fog computing resources and an energy-efficient passive optical network (PON)-based backhaul interconnects the VLC access points (APs). We investigate the joint optimization of the allocation of VLC and fog computing resources aiming to minimize processing and networking power consumption. We develop a mixed-integer linear programming (MILP) model that jointly optimizes access points (APs) selection, wavelength assignment and fog computing resources allocation to serve the requests of users. Compared to a statically preconfigured AP selection and wavelength assignment based on maximizing signal-to-interference-plus-noise ratio (SINR) independently of fog resource placement, the joint optimization achieves lower average total power consumption, driven by a reduction in processing power. These savings are attributed to the ability of the joint optimization to allow AP sharing across users and selecting more energy-efficient wavelengths that still satisfy quality of service (QoS) requirements.**

*Index Terms*— **energy efficient networks, fog computing, mixed integer linear programming (MILP), passive optical networks (PONs), visible light communication (VLC), resource allocation**

## I. INTRODUCTION

The massive proliferation of Internet-connected devices is placing unprecedented pressure on the radio frequency (RF) spectrum, leading to spectrum congestion that current wireless technologies struggle to address [1]. Visible light communication (VLC) has emerged as a compelling solution that exploits the vast unlicensed optical spectrum to deliver high-speed data transmission while simultaneously providing illumination. This dual functionality positions VLC as an energy efficient technology, particularly for indoor environments where users spend over 80% of their time [2], [3]. Moreover, VLC is increasingly recognized as a key enabling technology for realizing the 6G vision, enabling multi-terabit-per-second data rates connectivity in indoor environments [4].

The backhaul network connecting VLC access points (APs) within a building to each other and to external networks is critical to enable VLC systems. Several backhaul technologies have been considered for indoor VLC deployments, including power line communication (PLC) [5], Ethernet over Light (EoL) [6], wireless approaches [7], and passive optical networks (PONs) [8]. Among these, PONs are particularly advantageous due to their high data rates, cost-effectiveness, and energy efficiency due to the use of passive components. Moreover, PONs resides in the access networks, where fog computing can serve processing demands closer to end users, reducing both power consumption and latency. The optimization of fog resources allocation has been extensively studied, with energy consumption, latency, and cost as primary optimization objectives [9], [10], [11].

To tackle both energy consumption and latency of VLC backhaul networks, we proposed an energy-efficient PON-based backhaul connectivity for a VLC-Enabled indoor fog computing environment. In this architecture, the allocation of the VLC resources, i.e. APs and wavelength was preconfigured to maximize the signal-to-interference-plus-noise ratio (SINR), to ensure that the quality-of-service (QoS) requirements are met and the allocation of computing resources over the proposed architecture is optimized to minimize processing and networking power consumption.

Maximizing the SINR, which tends to favor high-power wavelengths [4], comes at the expense of energy efficiency. Assigning each user to a dedicated AP using the highest-power wavelength neglects the potential energy savings from sharing APs across users or selecting efficient wavelengths that can satisfy QoS requirements.

Furthermore, the static allocation of APs and wavelengths does not account for the impact that AP and wavelength assignment can have on the optimal placement of processing resources. Jointly optimizing AP, wavelength, and processing resource allocation can reduce the networking power consumption associated with a given processing node. Consequently, a processing node that appears less energy-efficient under a static allocation may become more energy-efficient than the nominally optimal node, leading to a lower overall energy consumption.

This paper therefore investigates the joint optimization of AP allocation, wavelength assignment, and fog computing resource allocation over the architecture we proposed in [12]. A mixed-integer linear programming (MILP) model is developed to jointly optimize AP allocation, wavelength assignment, and processing demand placement with the objective of minimizing the total networking and processing power consumption while satisfying the QoS requirements of all users. The proposed approach is evaluated against a baseline model in which AP selection and wavelength assignment are predetermined based on the maximum SINR criterion [4].

The remainder of this paper is organized as follows: Section II describes the system model, Section III presents the MILP formulation, Section IV discusses the results, and Section V concludes the paper.

## II. THE SYSTEM MODEL

The system architecture considered in this work is shown in Fig. 1. It consists of two main components: the in-building network and the fog and cloud computing resources.

### a. *In-building Network*

The in-building network encompasses the VLC systems deployed within rooms and the PON providing backhaul connectivity between rooms and to higher-tier fog and cloud resources [13]. The building consists of four rooms, each with eight VLC APs and eight user devices. VLC is employed to provide downlink connectivity from APs to user devices. VLC is typically unsuitable for uplink transmission, as directing light sources from user devices toward the ceiling can cause glare and discomfort [14]. Therefore, infrared (IR) communication is assumed for the uplink direction. This study focuses exclusively on the VLC downlink, while uplink optimization is left for future work.

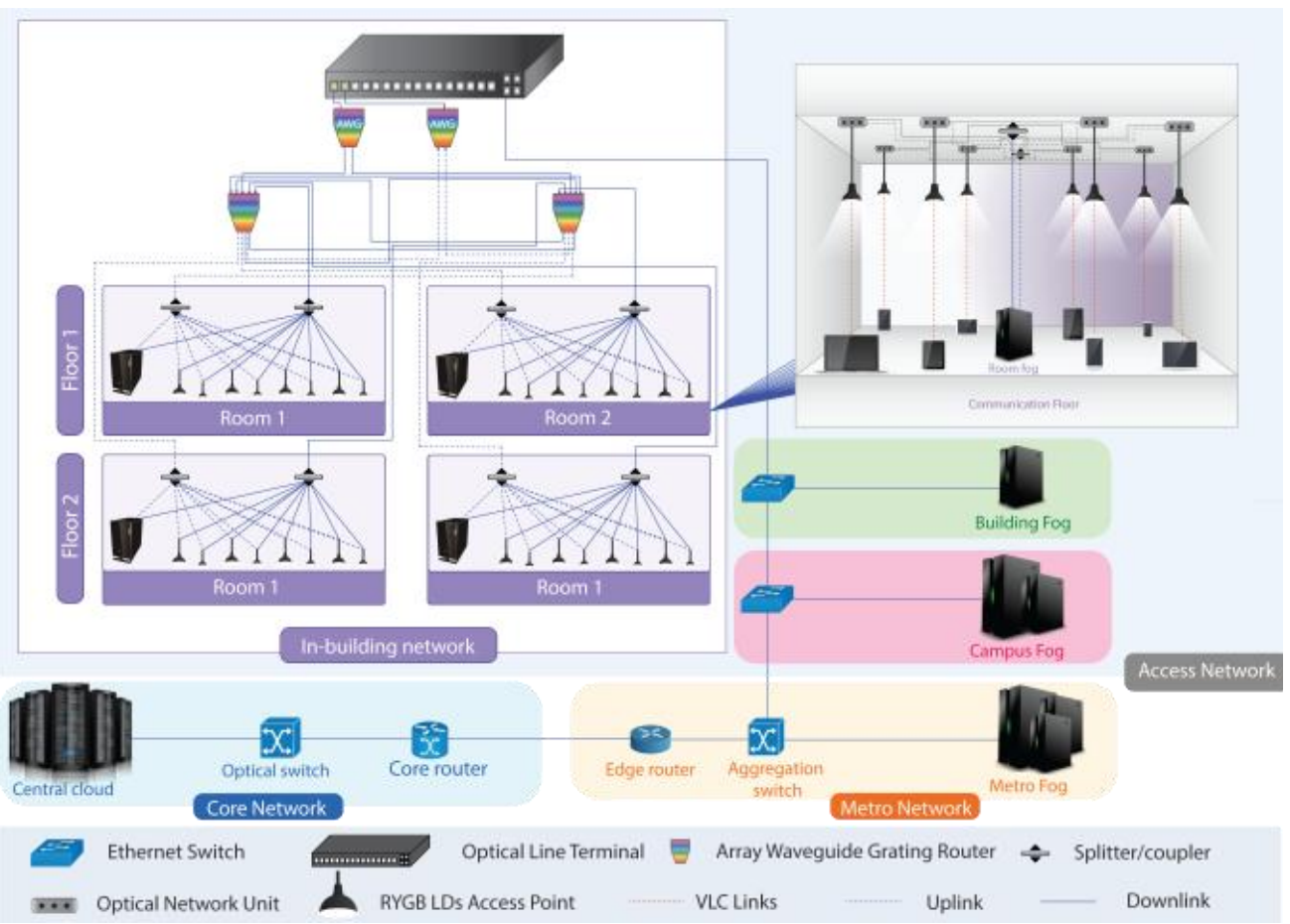

Fig. 1. The in-building PON-based VLC backhaul system with an end-to-end fog/cloud architecture.

Each VLC AP is composed of four RYGB laser diodes (LDs) mounted on the ceiling, functioning simultaneously as light sources and VLC transmitters. On the receiver side, user devices are equipped with angle diversity receivers (ADRs), where each branch can receive a WDM optical beam, enabling wavelength-selective reception from multiple directions.

The backhaul network is a time wavelength division multiplexing (TWDM)-PON, adopting the wavelength-routed architecture proposed in [14]. The proposed architecture extends conventional PON, originally designed for north-south traffic between users and the core network, to support east-west communication between users. The architecture is organized into four PON groups corresponding to the four rooms, all connected to an optical line terminal (OLT). Each AP is equipped with a tunable optical network unit (ONU) to support transmission and reception over multiple wavelengths. Within each room, a splitter divides the downstream optical signal to the APs, while a combiner aggregates upstream traffic from the APs. Two 5×5 arrayed waveguide grating routers (AWGRs) are employed to passively route traffic, enabling east-west communication between APs across rooms for accessing fog resources within the building, and north-south communication for accessing higher-tier fog and cloud resources. Two arrayed waveguide gratings (AWGs) serve as multiplexers and demultiplexers, connecting the in-building network to the OLT, which provides connectivity to the metro and core networks.

### *B. Fog-cloud Computing Architecture*

The fog computing resources are distributed across multiple tiers. Within the building, two types of processing resources are available: idle user devices that can offer their spare processing capacity, and dedicated room fog servers (RFSs) for each room. These in-building resources are complemented by higher-tier fog nodes, including a building fog server (BFS), a campus fog server (CFS), and a metro fog server (MFS), with the central cloud serving as the final processing tier. In this hierarchical structure, processing demands can be offloaded to tiers closer to the user, thereby reducing latency and networking power consumption. Fog nodes closer to the user offer lower networking power consumption but have constrained capacity and higher energy per operation. Conversely, the cloud offers greater processing capacity and lower energy per operation but is associated with significantly higher networking power consumption and higher idle power consumption. Therefore, the optimal placement of processing demands in this hierarchical architecture involves a tradeoff between processing and networking power consumption.

## III. A MILP Model for Joint Optimization of VLC and Fog Resources

This section extends the MILP model presented in [12] to jointly optimize VLC resources allocation and processing demands placement to minimize the total networking and processing power consumption. As explained above, the joint optimization can reduce the networking power consumption associated with a given processing node leading to a lower overall energy consumption compared to an architecture where the allocation of APs and wavelengths is preconfigured to maximize SINR. The model is extended to incorporate the selection of APs and wavelengths connecting user devices offering their processing capacity as optimization variables, rather than assuming these are preconfigured inputs. In the following we present the set, parameters, variables used in the extension of the model in [12]. For the complete mathematical formulation of the original model, the reader is referred to [12].

**Sets:**

| | |
|---|---|
| $N$ | Set of all nodes. |
| $N_i$ | Set of neighbors of node $i$, $i \in N$. |
| $M$ | Set of all user devices. |
| $P$ | Set of processing nodes including the user devices, RFSs, BFS, CFS, MFS, and cloud server. |
| $UD$ | Set of user devices generating demands. |
| $AP$ | Set of APs connected to source nodes. |
| $CP$ | Set of APs connected to processing user devices. |
| $CO$ | Set of couplers. |
| $SP$ | Set of splitters. |
| $Lo$ | Set of user device locations within each room. |
| $Br$ | Set of branches of an ADR. |
| $V$ | Set of the wavelengths generating the shade of white $V = \{R, Y, G, B\}$. |
| $W$ | Set of wavelengths used in PON backhaul network. |

**Parameters:**

| | |
|---|---|
| $PR$ | Maximum power consumption of an AP when serving a user device using the red wavelength. |
| $PY$ | Maximum power consumption of an AP when serving a user device using the yellow wavelength. |
| $O_{ur}^{av}$ | Received optical power by user device $u$ using receiver branch $r$ and wavelength $v$ from AP $a$, where $a \in AP, u \in M, r \in Br, v \in V$. |
| $S$ | Preamplifier noise. |
| $K_{ur}^{av}$ | Background noise due to unmodulated optical power received by user device $u$ using receiver branch $r$ and wavelength $v$ of AP $a$, where $a \in AP, u \in M, r \in Br, v \in V$. |
| $Y$ | Minimum SINR threshold to satisfy QoS constraint. |
| $Z$ | A very large number. |

**Variables:**

| | |
|---|---|
| $\lambda_{ijw}^{ud}$ | Traffic flow between source and destination pair $(u, d)$ traversing the physical link $(i, j)$ using wavelength $w$, where $u \in UD$, $d \in P, i \in N, j \in N_i, w \in W$. |
| $\Theta_a$ | Binary variable, $\Theta_a = 1$, if AP $a$ is activated, otherwise, $\Theta_a = 0, a \in AP \cup CP$. |
| $R_a$ | Binary variable, $R_a = 1$, if AP $a$ serves processing user devices using the red wavelength, otherwise, $R_a = 0, a \in CP$. |
| $Y_a$ | Binary variable, $Y_a = 1$, if AP $a$ serves processing user devices using |

the yellow wavelength, otherwise, $Y_a = 0, a \in CP$.

$GB_a$ Binary variable, $GB_a = 1$, if AP $a$ serves processing user devices using the green and blue wavelengths, otherwise, $GB_a = 0, a \in CP$.

$\text{ɥ}_{ur}^{av}$ SINR at user device $u$ , assigned to AP $a$, using a receiver branch $r$ and wavelength $v$, where $u \in M, a \in N_i, r \in Br, v \in V$.

$\tau_{ur}^{av}$ Binary variable, $\tau_{u,r}^{a,v} = 1$, if user device $u$ is assigned to AP $a$ using a receiver branch $r$ and wavelength $v$, otherwise $\tau_{u,r}^{a,v} = 0$, where $u \in M, a \in AP, r \in Br, v \in V$.

$\text{ʑ}_{ua}$ Binary variable, $\text{ʑ}_{ua} = 1$, if user device $u$ is assigned to AP $a$, otherwise, $\text{ʑ}_{ua} = 0$, where $u \in M, a \in N_i$.

$\psi_{urm}^{abv}$ Auxiliary continuous variable representing the product of the SINR, $\text{ɥ}_{ur}^{av}$, and the binary allocation variable $\tau_{mr}^{bv}$, where $u, m \in M, a, b \in AP, r \in Br, v \in V, u \neq m, a \neq b$.

$X_{ijw}^{sd}$ Auxiliary continuous variable representing the product of the traffic flow variable $\lambda_{ijw}^{ud}$ and the binary AP assignment variable $\text{ʑ}_{ua}$, where $s \in UD, d \in P, i \in M, j \in Ni, w \in W, s \neq d$.

The objective of the model is to minimize the total power consumption:

$$minimize\, TPC = PC + PN, \quad (1)$$

where $TPC$ is the total power consumption composed of the processing power consumption $PC$, and the networking power consumption $PN$. $PC$ and $PN$ are calculated in [12], with the power consumption of the APs (Equations (10)-(15) in [12]) replaced by the following formulation:

$$PA = \sum_{a \in AP} \Theta_a PR + \sum_{a \in CP} R_a PR + Y_a PY + GB_a PGB. \quad (2)$$

Each AP consumes fixed power for each wavelength it actively transmits, independently of the carried traffic load, as the LD output power required for transmission and illumination does not scale with the bit rate. The APs serving source user devices are preconfigured to use the red wavelength and therefore consume $PR$ when activated, while the APs serving processing user devices consume the power of each wavelength assigned by the model additively.

The flow conservation, capacity and processing allocation constraints from [12] are retained. In the following we present the additional constraints introduced to extend the model.

### A. SINR calculations and QoS Constraint

$$\text{ɥ}_{ur}^{av} = \frac{O_{ur}^{av}\tau_{ur}^{av}}{\sum_{m \in M, m \neq u}\sum_{b \in AP, a \neq b}\sum_{r \in Br}(O_{ur}^{bv} - K_{ur}^{bv})\ \tau_{ur}^{av,} + \sum_{b \in AP, a \neq b} \text{ɥ}_{ur}^{av} K_{ur}^{bv} + S} \quad (3)$$
$$\forall u \in M, a \in AP, r \in Br, v \in V.$$

Equation (3) calculates the SINR at each assigned user device as the ratio of the received signal power to the aggregate interference from other APs and noise.

$$\sum_{m \in M, m \neq u}\sum_{b \in AP, a \neq b}\sum_{r \in Br}(O_{ur}^{bv} - K_{ur}^{bv})\,\text{ɥ}_{ur}^{av}\tau_{mr}^{bv} + S\text{ɥ}_{ur}^{av} + \sum_{b \in AP, a \neq b}\text{ɥ}_{ur}^{av}K_{ur}^{bv} = O_{ur}^{av}\tau_{ur}^{av}, \quad (4)$$
$$\forall u \in M, a \in AP, r \in Br, v \in V.$$

Constraint (4)

(4) linearizes equation (3) by rearranging the SINR expression, as aggregate interference and noise terms are on the left-hand side and the desired signal power is on the right-hand side [4].

$$\psi_{urm}^{abv} = \text{ɥ}_{ur}^{av}\tau_{mr}^{bv}, \forall u, m \in M, a, b \in AP, r \in Br, v \in V, u \neq m, a \neq b. \quad (5)$$

Equation (5) introduces the auxiliary variable $\psi_{urm}^{abv}$ as the product of the SINR variable $\text{ɥ}_{ur}^{av}$ and the binary variable $\tau_{mr}^{bv}$.

$$\psi_{urm}^{abv} \leq Z\tau_{mr}^{bv}, \quad (6)$$
$$\forall u, m \in M, a, b \in AP, r \in Br, v \in V, u \neq m, a \neq b.$$

$$\psi_{urm}^{abv} \leq \text{ɥ}_{ur}^{av}, \quad (7)$$
$$\forall u, m \in M, a, b \in AP, r \in Br, v \in V, u \neq m, a \neq b.$$

$$\psi_{urm}^{abv} \geq \text{ɥ}_{ur}^{av} - (1 - \tau_{mr}^{bv})Z, \quad (8)$$
$$\forall u, m \in M, a, b \in AP, r \in Br, v \in V, u \neq m, a \neq b.$$

$$\psi_{urm}^{abv} \geq 0, \quad (9)$$
$$\forall u, m \in M, a, b \in AP, r \in Br, v \in V, u \neq m, a \neq b.$$

Constraints (6)-(9) are the linearization constraints for equation (5).

$$\text{ɥ}_{ur}^{av} \geq Y\,\tau_{ur}^{av}, \forall u \in M, a \in AP, r \in Br, v \in V. \quad (10)$$

Constraint (10) ensures that the SINR resulting from the allocation of resources at each user device meets the minimum QoS threshold.

### B. VLC Resource Allocation Constraints

$$\sum_{u \in M}\sum_{r \in Br}\tau_{ur}^{av} \leq 1, \forall a \in AP, v \in V. \quad (11)$$

Constraint (11) ensures that each wavelength of an AP branch is allocated to at most one user device at a time.

$$\sum_{a \in AP}\sum_{r \in Br}\sum_{v \in V}\tau_{ur}^{av} \leq 1, \forall u \in M. \quad (12)$$

Constraint (12) ensures that each user device is assigned to at most one AP using a single receiver branch and a single wavelength.

$$Z\sum_{r \in Br}\sum_{v \in V}\tau_{ur}^{av} \geq \text{ʑ}_{ua}, \forall u \in M, a \in \text{AP}. \quad (13)$$

$$\sum_{r \in Br}\sum_{v \in V}\tau_{ur}^{av} \leq Z\,\text{ʑ}_{ua}, \forall u \in M, a \in AP. \quad (14)$$

Constraints (13) and (14) relate the continuous allocation variable $\tau_{ur}^{av}$ to its binary equivalent $\text{ʑ}_{ua}$.

$$R_a = \sum_{u \in M}\sum_{r \in Br}\tau_{ur}^{aR}, \forall a \in CP. \quad (15)$$

$$Y_a = \sum_{u \in M}\sum_{r \in Br}\tau_{ur}^{aY}, \forall a \in CP. \quad (16)$$

$$GB_a = \sum_{u \in M}\sum_{r \in Br}\tau_{ur}^{aG,B}, \forall a \in CP. \quad (17)$$

Constraints (15), (16) and (17) relate the wavelength indicators $R_a$, $Y_a$ and $GB_a$ to the allocation variable $\tau_{ur}^{av}$.

### C. Traffic Routing Constraints

$$X_{ijw}^{sd} = \lambda_{ijw}^{sd}\text{ʑ}_{ia}, \forall s \in UD, d\ in\ P, i \in M, w \in W, s \neq d. \quad (18)$$

Constraint (18) ensures that traffic from a user device is only routed through its assigned AP. Constraint (18) is nonlinear. It is replaced by the following linear inequalities:

$$X_{ijw}^{sd} \leq Z\,\text{ʑ}_{ia}, \forall s \in UD, d\ in\ P, i \in M, w \in W, s \neq d. \quad (19)$$

$$X_{ijw}^{sd} \geq \lambda_{ijw}^{sd} - (1 - \text{ʑ}_{ia})Z, \forall s \in UD, d\ in\ P, i \in M, w \in W, s \neq d. \quad (20)$$

$$\sum_{j \in AP \cap N_i}\lambda_{ijw}^{sd} \leq 0, \forall s \in \text{UD}, d \in P, i \in CO, w \in W. \quad (21)$$

Constraint (21) prevents traffic from being forwarded from couplers toward APs, ensuring that traffic flows only in the upstream direction from APs to couplers.

$$\sum_{j \in SP \cap N_i}\lambda_{ijw}^{sd} \leq 0, \forall s \in \text{UD}, d \in P, i \in AP, w \in W. \quad (22)$$

Constraint (22) prevents traffic from being forwarded from APs toward splitters, ensuring that traffic flows only in the downstream direction from splitters to APs.

## IV. PERFORMANCE EVALUATION

We evaluated the improvement in energy efficiency obtained by jointly optimizing the allocation of VLC and processing resources to minimize the total power consumption of serving requests in the fog-cloud architecture, referred to as optimized VLC and processing (OVP) against optimizing the processing resources allocation while preconfiguring the AP selection and wavelength assignment based on SINR maximization, referred to as the optimized processing (OP) model [4].

Both models consider fixed user locations within the room and enforce a minimum SINR threshold of 12 dB to satisfy the QoS requirement [12]. Two user distribution cases are considered: a clustered case where user devices are clustered around four APs, as shown in Fig. 2, and a dispersed case where each user device is positioned closer to a different AP, as shown in Fig. 3. In the clustered case, user devices are grouped in pairs, with each pair clustered around a single AP. User devices sharing the same AP and receiver branch are assigned different wavelengths to avoid interference. In this case, alternative assignment would violate the QoS constraint. In the dispersed case, each user device is assigned to a dedicated AP using the red wavelength, the wavelength that provides the highest SINR, which is the preconfigured assignment used in the OP model based on [4]. However, as shown in Table 1, the red wavelength is the least energy-efficient among the available wavelengths, and assigning each user to a dedicated AP forgoes potential power savings from sharing APs across multiple users. Consequently, only the dispersed case is considered in this work, as it provides the necessary flexibility to observe the impact of joint optimization.

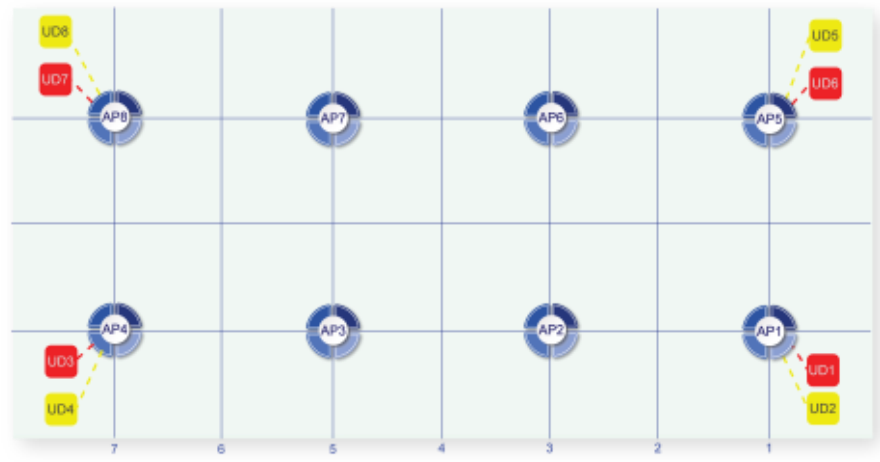

Fig. 2. Top view of the room layout showing user device locations and static AP assignments for the clustered distribution case.

User devices generate requests, each with a processing demand range from 6 GFLOPs to 20 GFLOPs and a networking demand from 0.3 Gbps to 1 Gbps. Device parameters are provided in Table 1 and Table 2 [12]. Since uplink optimization is out of scope for this work, as discussed in Section II, the power consumption of user devices generating demands is not considered in either model. The comparison is therefore based on the downlink power consumption resulting from processing placement decisions.

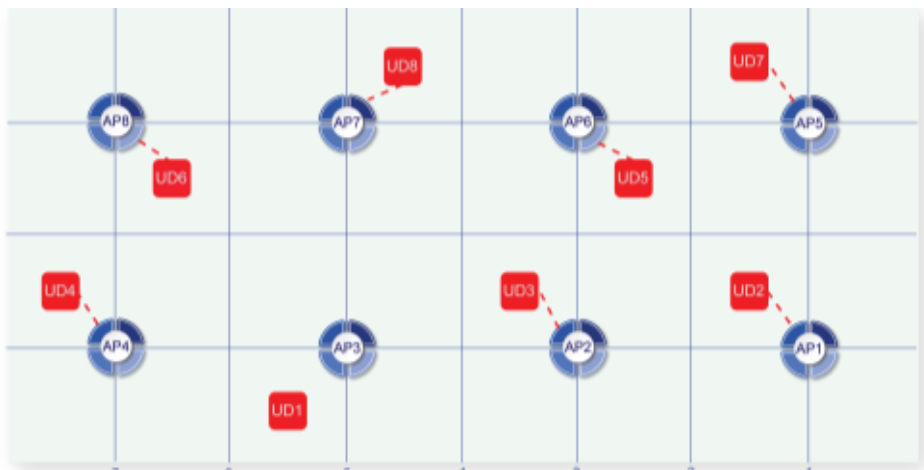

Fig. 3. Top view of the room layout showing user device locations and static AP assignments for the dispersed distribution case.

TABLE 1: NETWORKING DEVICES PARAMETERS VALUES [12] .

| Network device | | Max power (W) | Idle power (W) | Capacity (GFLOPs) | Efficiency (W/ GFLOPs) |
|---|---|---|---|---|---|
| AP | Red | 7.2 | 4.32 | 2.5 | 1.52 |
| | Yellow | 4.5 | 2.7 | 2.5 | 0.72 |
| | Green | 2.7 | 1.62 | 2.5 | 0.432 |
| | Blue | 2.7 | 1.62 | 2.25 | 0.485 |
| ONU | | 15 | 9 | 10 | 0.9 |
| Ethernet switch | | 300 | 180 | 160 | 1.125 |
| Aggregation Sw. | | 435 | 261 | 240 | 0.725 |
| Edge router | | 435 | 261 | 240 | 0.725 |
| Optical switch | | 750 | 450 | 480 | 0.625 |
| Core router | | 344 | 206.4 | 3200 | 0.043 |

TABLE 2: PROCESSING DEVICES PARAMETERS VALUES. [12]

| **Processing node** | **Max power (W)** | **Idle power (W)** | **Capacity (GFLOPs)** | **Efficiency (W/ GFLOPs)** |
|---|---|---|---|---|
| Cloud server | 1100 | 660 | 1612.8 | 0.27 |
| MFS | 750 | 450 | 403.2 | 0.74 |
| CFS | 350 | 210 | 121.6 | 1.15 |
| BFS | 305 | 183 | 99 | 1.23 |
| RFS | 65 | 39 | 64 | 0.41 |
| User devices | 18 | 10.8 | 12.288 | 0.55 |

### A. *Scenario 1: Requests from Multiple Rooms*

This scenario considers two user devices per room generating requests, while the remaining user devices offer their processing capabilities. Fig. 4 shows the optimal processing workload allocation for the OVP and OP models. In most cases, the allocation results are identical across both models. However, subtle differences arise at 9 GFLOP and 10 GFLOPs.

At 9 GFLOPs, both models activate the same processing nodes, a user device and an RFS. However, as shown in Fig. 5, the OVP model assigns UD7 in Floor 1 Room 2 to AP5 using the yellow wavelength, rather than the red wavelength used in the OP model. This reflects the ability of the OVP model to select a more energy-efficient wavelength while still satisfying the QoS constraint.

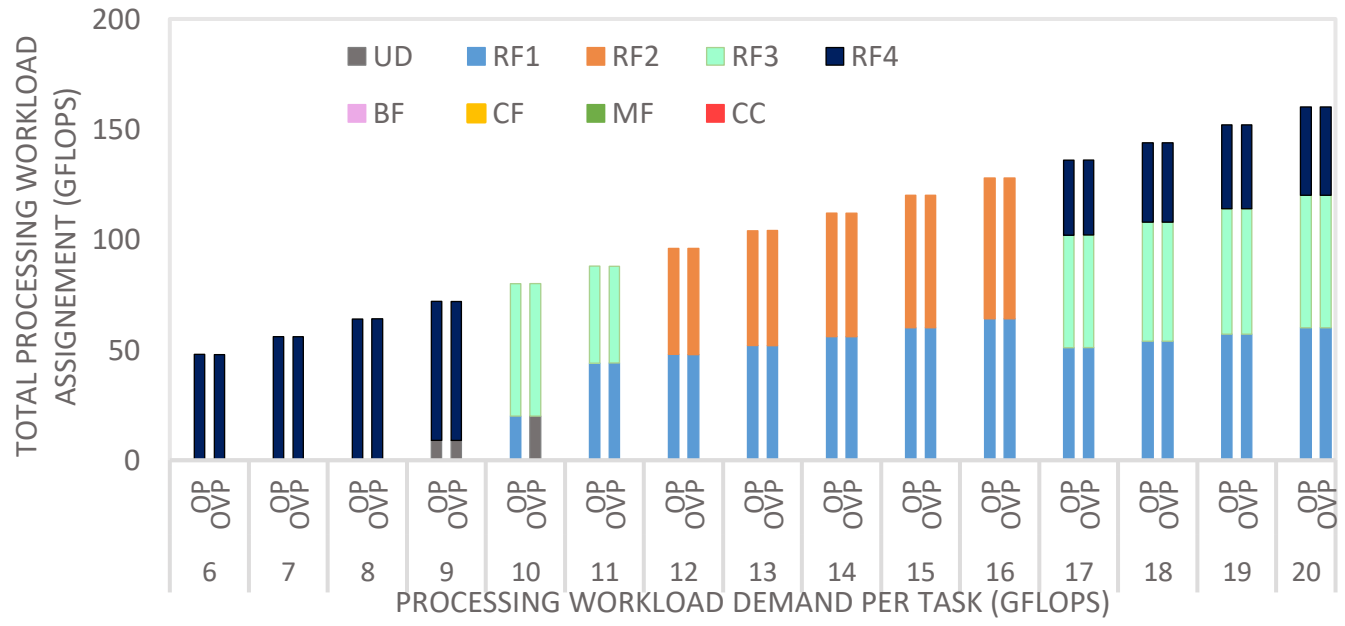

Fig. 4. Optimal processing workload allocation for Scenario 1.

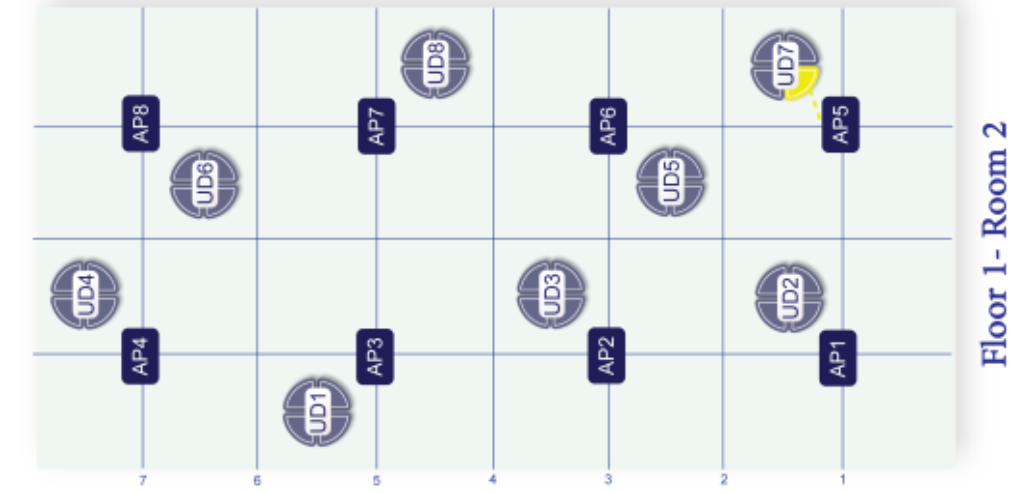

Fig. 5. Optimal AP and wavelength assignment for Scenario 1 at 9 GFLOPs.

At 10 GFLOPs, the two models diverge in their processing placement decisions. The OP model activates two RFSs (RFS1 and RFS3), while the OVP model activates two user devices and RF3. As shown in Fig. 6, under the OVP model, UD2 and UD7 in Floor 1 Room 2 are both assigned to AP5 using the red and yellow wavelength, respectively. The red wavelength, although the least energy-efficient, provides the highest SINR and is therefore assigned to UD2, the more distant of the two devices, to satisfy the SINR threshold; the closer device, UD7, is instead assigned the more energy-efficient yellow wavelength, which remains sufficient given its shorter distance from AP5. Placing the demand on two user devices served by a single AP requires the activation of an ONU and two wavelengths, resulting in higher networking power consumption than placing the demand on an RFS accessed through a single ONU. However, this increase is outweighed by the reduction of processing power consumption, leading to a lower overall power consumption.

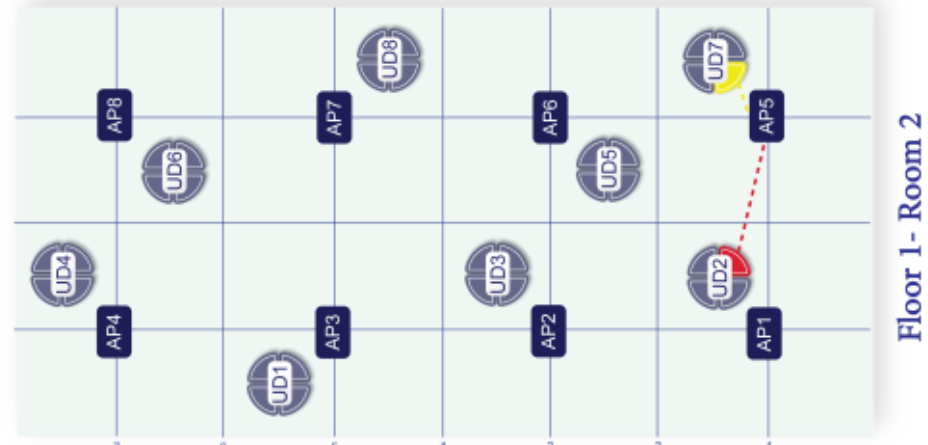

Fig. 6. Optimal AP and wavelength assignment for Scenario 1 at 10 GFLOPs.

Fig. 7 presents the power consumption comparison between the two models. As shown in Fig. 7a, at 9 GFLOPs, processing power consumption remains unchanged between the two models, as the processing allocation results are identical in both cases. At 10 GFLOPs, the OVP model achieves a maximum 12% reduction in processing power consumption by serving two of the demands using two user devices instead of an additional room fog server, avoiding its higher processing power cost. Fig. 7b shows that at 9 GFLOPs the OVP model achieves a maximum 2% reduction in networking power consumption. At 10 GFLOPs, however, networking power consumption increases, since reaching the room fog server only requires powering the ONU, whereas reaching user devices also require powering the AP serving it over the red and yellow wavelengths. Fig. 7c shows a maximum total power saving of 1% at both 9 and 10 GFLOPs.

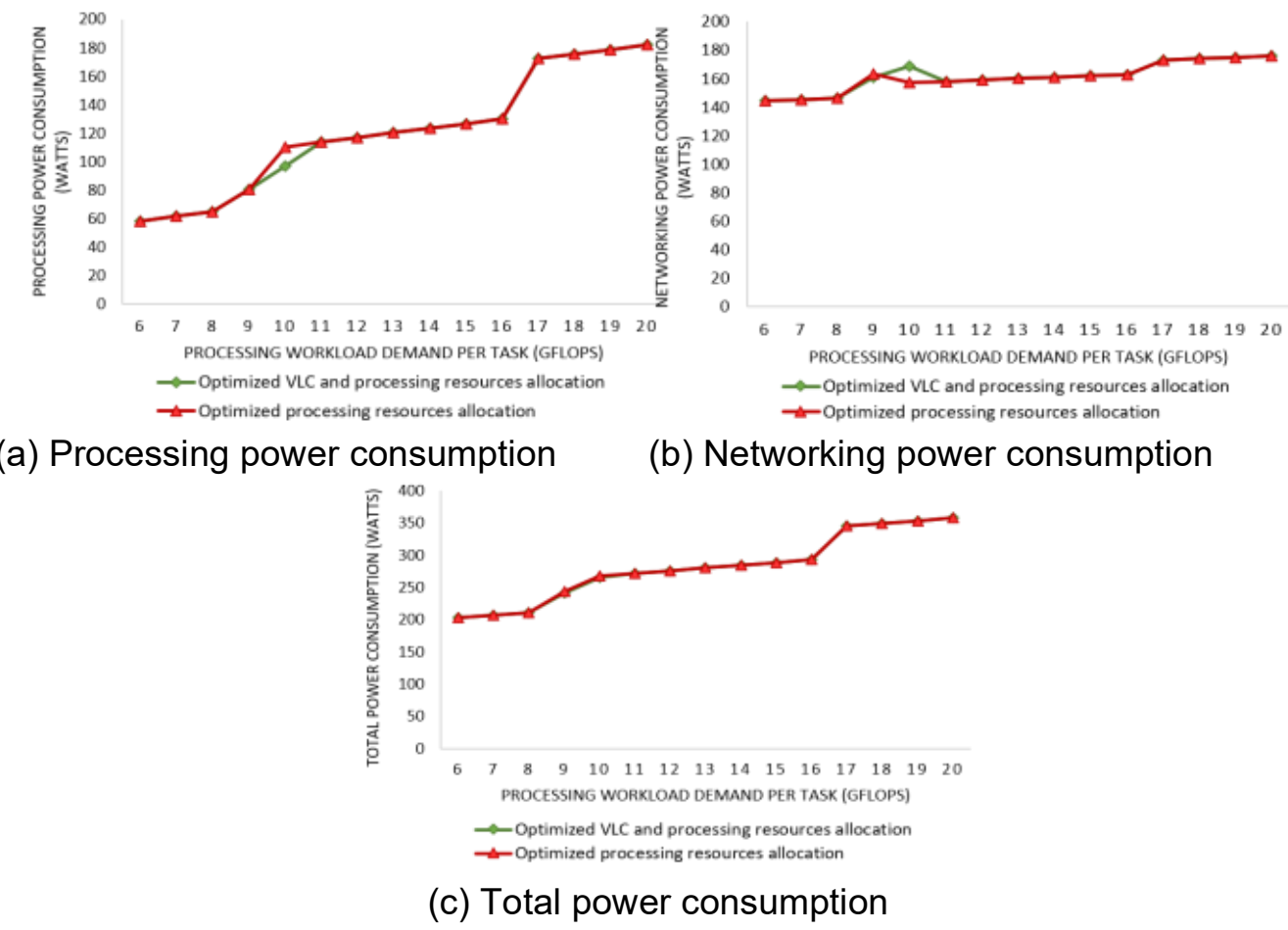

Fig. 7: Power consumption comparison between the OP and OVP models for Scenario 1.

*B. Scenario 2: Requests from a Single Room*

In this scenario, all eight user devices in one room generate demands, while the user devices in other rooms offer their idle processing resources.

Fig. 8 shows the optimal processing placement under both models. At 6 GFLOPs, both models allocate tasks to user devices, though the selection differs, as discussed in the following subsection. From 7 GFLOPs onwards, the two models diverge. At 7 GFLOPs, the OP model activates three RFSs (RFS1, RFS2, and RFS3), while the OVP model allocates tasks to two RFSs and two user devices sharing a single AP, avoiding the cost of activating an additional RFS. From 8 to 11 GFLOPs, the OP model activates all RFSs, which remains underutilized due to network bottlenecks, while the OVP model activates user devices. At 12 GFLOPs, both models allocate tasks to user devices, with differences attributable to AP selection as discussed below. Beyond 12 GFLOPs, the BF server is activated alongside RFS1 as in-building resources become insufficient, with an additional RFS required from 16 GFLOPs onwards.

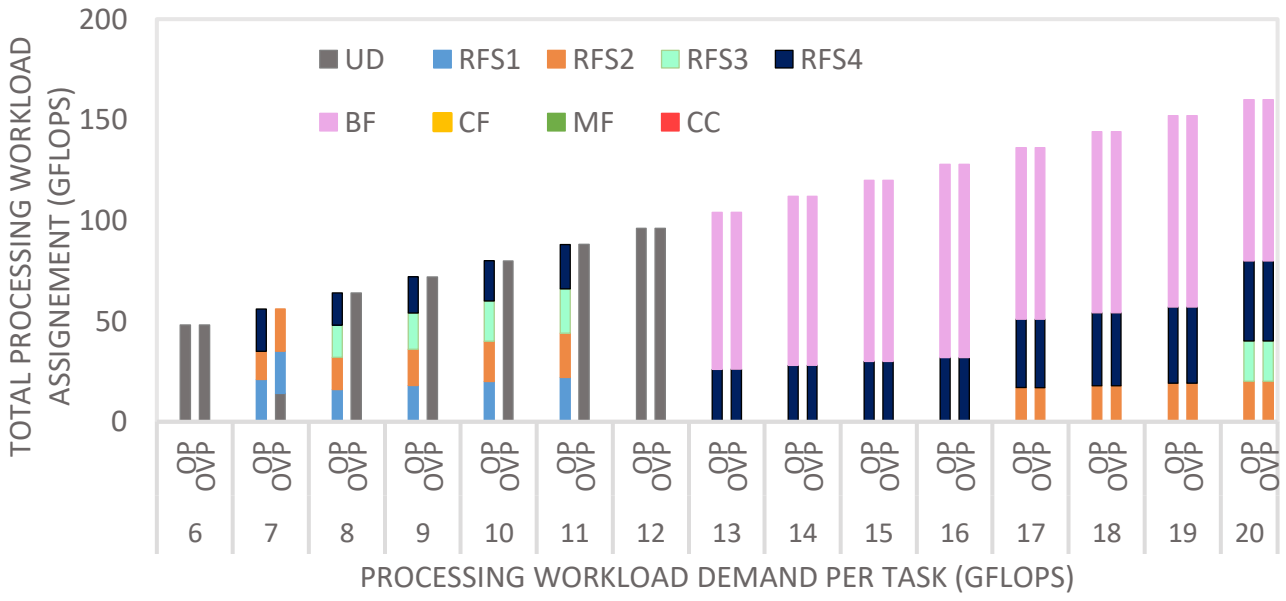

Fig. 8. The optimal processing allocation for the second scenario.

Fig. 9 shows the OVP model AP and wavelength assignments at 6 GFLOPs, where four user devices are activated across two rooms. Each processing user device is assigned to a dedicated AP using the yellow wavelength rather than the red wavelength used in the OP model. Since the yellow wavelength consumes less power per bit while still satisfying the SINR threshold, this selection reduces networking power consumption relative to the OP model.

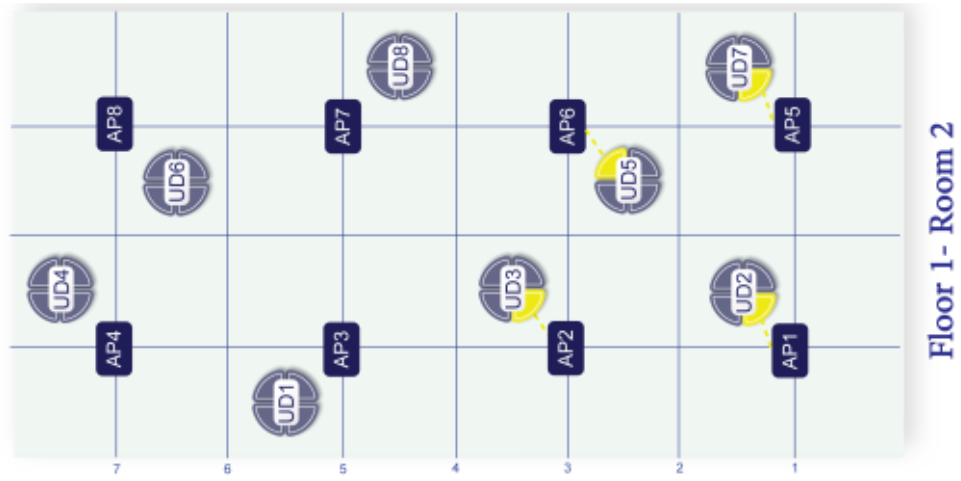

Fig. 9. Optimal AP and wavelength assignment for Scenario 2 at 6 GFLOPs.

At 7 GFLOPs, a similar assignment to the one shown in Fig. 6 is observed, where two user devices share a single AP using the red and yellow wavelength.

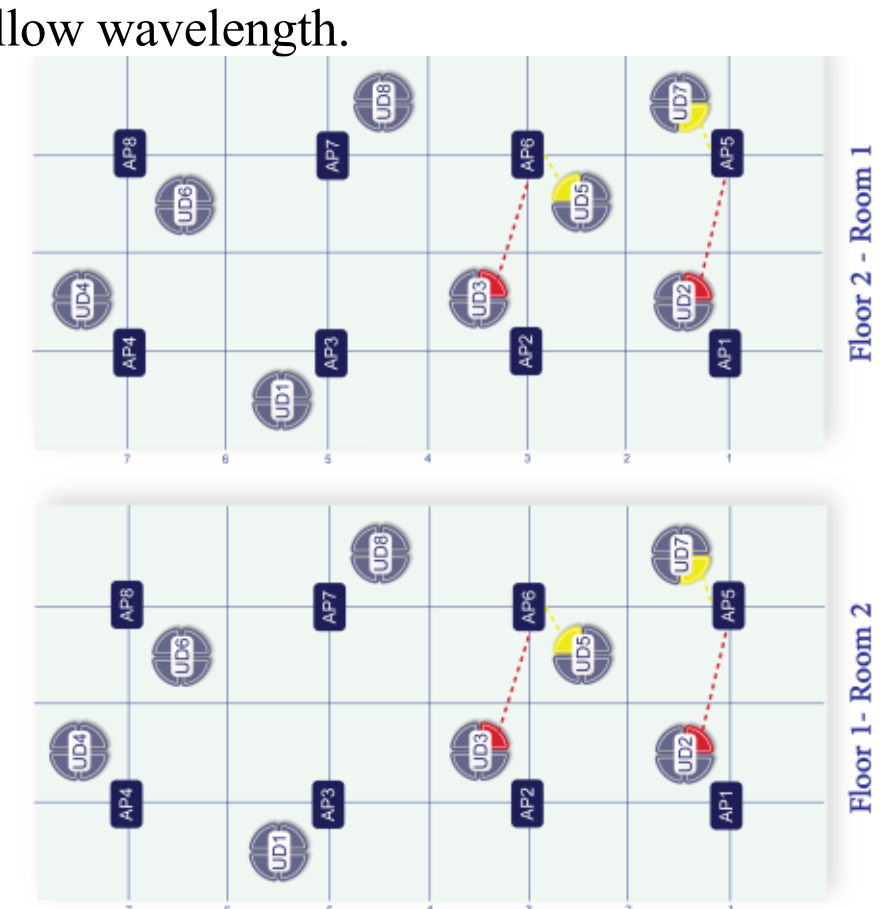

Fig. 10. Optimal AP and wavelength assignment for Scenario 2 at 8-11 GFLOPs.

From 8 to 11 GFLOPs, the assignment results are presented in Fig. 10. At this demand range, each user device's processing capacity (12.8 GFLOPs) is sufficient to serve only a single task. Eight user devices are therefore activated, one per task. The networking demand per task in this range (0.4-0.5 Gbps) remains within the AP capacity of 1.1 Gbps for two tasks combined, allowing a single AP to serve two user devices. In each room, two APs are activated, with each AP serving a nearby and a distant user device using the yellow and red wavelengths, respectively.

The selection of these specific user devices is not unique, any other pair of user devices that satisfies the QoS constraint connecting to an AP could achieve the same result, since the passive optical network ensures equal networking cost across all rooms.

At 12 GFLOPs, the assignment results are presented in Fig. 11. Each task requires 0.6 Gbps, and since the AP capacity is 1.1 Gbps, two tasks together (1.2 Gbps) exceed what a single AP can support (1.1 GFLOPs). The model therefore assigns each user device to a dedicated AP using the yellow wavelength.

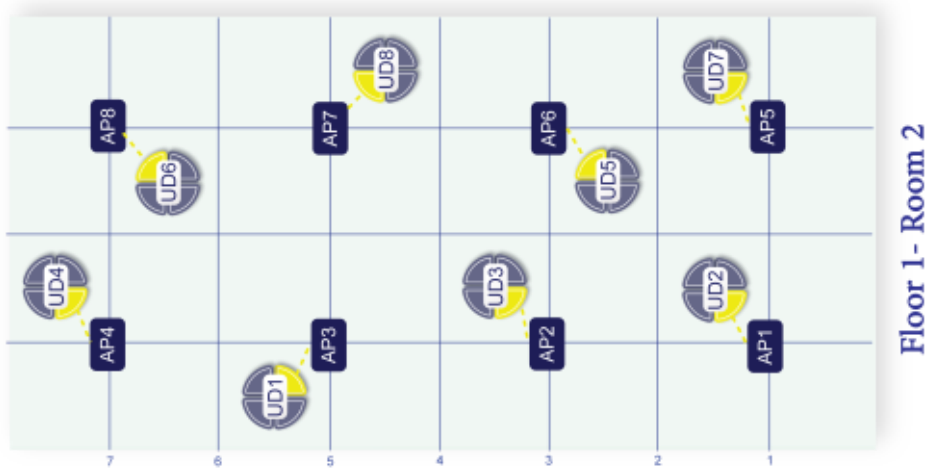

Fig. 11. Optimal AP and wavelength assignment for Scenario 2 at 12 GFLOPs.

Fig. 12 presents the power consumption comparison between the two models for Scenario 2. As shown in Fig. 12a, the OVP model achieves a maximum processing power consumption reduction of 32%, due to the activation of user devices instead of RFSs, with reductions observed between 7 and 11 GFLOPs and no difference between the two models at 6 and 12 GFLOPs. Fig. 12b shows that networking power consumption increases for most of the considered demand range (7-11 GFLOPs), since reaching a room fog server only requires activating the ONU connected to it, whereas reaching a user device also requires powering the AP serving it over the red and yellow wavelengths; even though a single AP is mostly shared across two user devices, this still results in higher networking power consumption than serving the same demand at a room fog server. A maximum networking power reduction of up to 8% is observed only at demand levels where user devices are selected for processing in both models (6 and 12 GFLOPs), at which point the OVP model's AP and wavelength selection alone accounts for the networking power difference. Fig. 12c shows that the OVP model still achieves a maximum total power saving of 5%, driven by the processing power reduction outweighing the networking power increase. Beyond 12 GFLOPs, both models yield identical processing and networking power consumption as the allocation decisions converge.

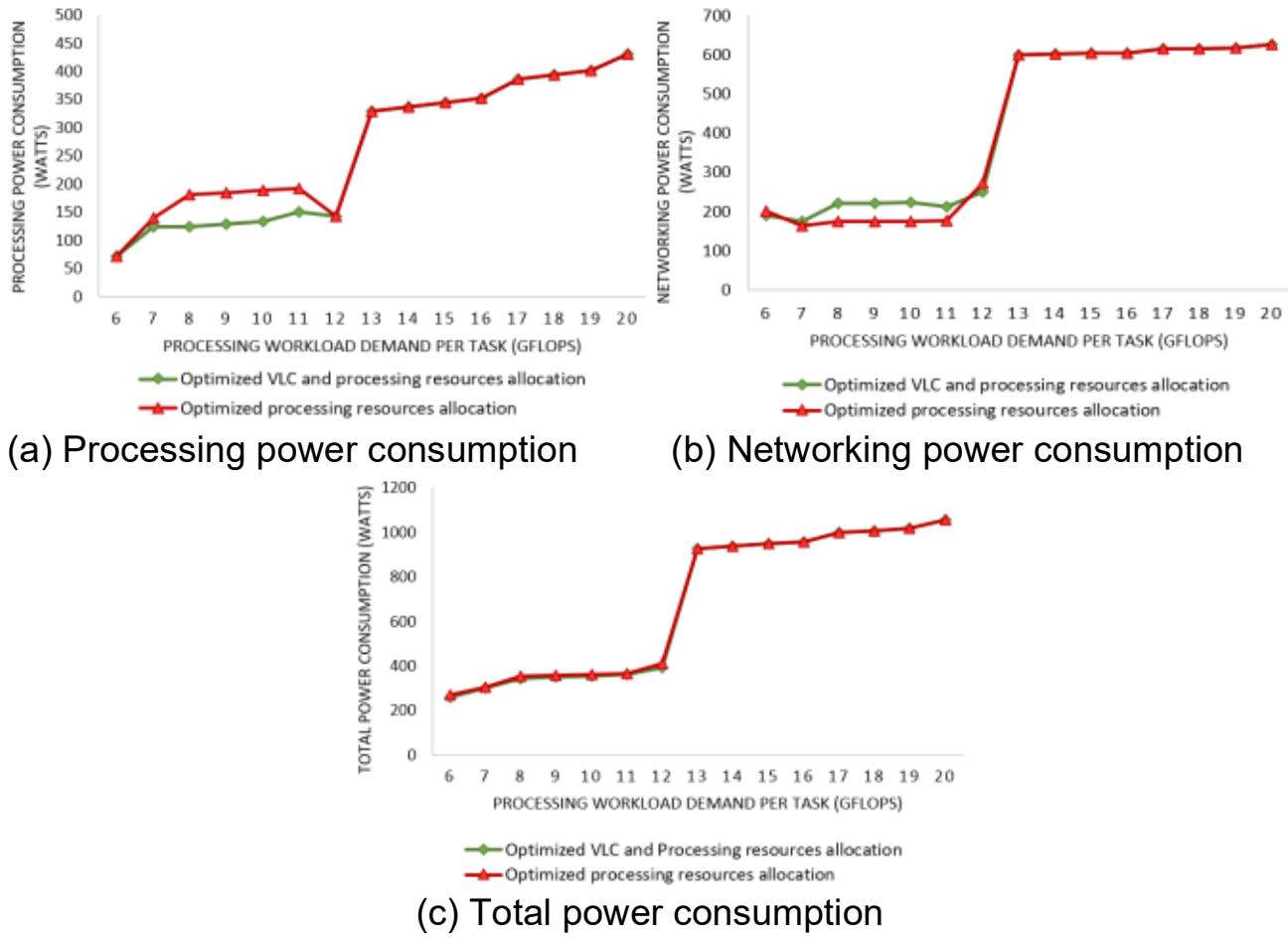

Fig. 12. Power consumption comparison between the OP and OVP models for Scenario 2.

## V. Conclusion

This paper investigated the joint optimization of VLC and fog computing resource allocation over a PON-based backhaul architecture that interconnects VLC APs and fog computing resources. A MILP model was developed to jointly optimize AP selection, wavelength assignment, and processing demand placement, with the objective of minimizing total networking and processing consumption while satisfying QoS constraints. The results demonstrated that the joint optimization of communication and computation achieved lower total power consumption compared to a baseline model where APs and wavelengths are preconfigured based on maximum SINR, driven by a substantial reduction in processing power consumption. The energy savings achieved are primarily due to more efficient processing resource placement, enabled by flexible AP sharing among users and the selection of energy-efficient wavelengths without compromising QoS [12].